\documentclass[final]{svjour3}
\smartqed  

\AtBeginDocument{
\setlength{\oddsidemargin}{\dimexpr(\paperwidth-\textwidth)/2-1in}
\setlength{\evensidemargin}{\oddsidemargin}
\setlength{\topmargin}{\dimexpr(\paperheight-\textheight)/2-\headheight-\headsep-1in}}

\usepackage{graphicx}
\usepackage{amssymb,amsmath}
\usepackage[latin9]{inputenc}
\usepackage{float}
\usepackage{multicol}

\begin{document}

\title{The concept of laser-based conversion electron M\"ossbauer spectroscopy for a precise energy determination of $^{229\mathrm{m}}$Th}

\author{Lars C. von der Wense \and
        Benedict Seiferle \and
        Christian Schneider \and
        Justin Jeet \and
        Ines Amersdorffer \and
        Nicolas Arlt \and
        Florian Zacherl\and
        Raphael Haas \and
        Dennis Renisch \and
        Patrick Mosel \and
        Philip Mosel \and
        Milutin Kovacev \and
        Uwe Morgner \and
        Christoph E. D\"ullmann \and
        Eric R. Hudson \and
        Peter G. Thirolf
}

\institute{L. C. von der Wense \and B. Seiferle \and I. Amersdorffer \and N. Arlt \and F. Zacherl \and P. G. Thirolf\at
              Ludwig-Maximilians-Universit\"at M\"unchen, 85748 Garching, Germany
           \and
           Ch. Schneider \and J. Jeet \and E. R. Hudson \at
           University of California, Los Angeles, California 90095, USA
            \and
           R. Haas$^{1,2,3}$ \and D. Renisch$^{1,2}$ \and Ch. E. D\"ullmann$^{1,2,3}$ \at
           $^1$ Johannes Gutenberg-Universit\"at Mainz, 55099 Mainz, Germany\\
           $^2$ Helmholtz-Institut Mainz, 55099 Mainz, Germany\\
           $^3$ GSI Helmholtzzentrum f\"ur Schwerionenforschung GmbH, 64291 Darmstadt, Germany
           \and
           Pa. Mosel \and Ph. Mosel \and M. Kovacev \and U. Morgner \at
           Leibniz Universit\"at Hannover, 30167 Hannover, Germany
}

\date{}

\maketitle
\begin{multicols}{2}
\begin{abstract}
$^{229}$Th is the only nucleus currently under investigation for the development of a nuclear optical clock (NOC) of ultra-high accuracy. The insufficient knowledge of the first nuclear excitation energy of $^{229}$Th has so far hindered direct nuclear laser spectroscopy of thorium ions and thus the development of a NOC. Here, a nuclear laser excitation scheme is detailed, which makes use of thorium atoms instead of ions. This concept, besides potentially leading to the first nuclear laser spectroscopy, would determine the isomeric energy to 40~$\mu$eV resolution, corresponding to 10~GHz, which is a $10^4$ times improvement compared to the current best energy constraint. This would determine the nuclear isomeric energy to a sufficient accuracy to allow for nuclear laser spectroscopy of individual thorium ions in a Paul trap and thus the development of a single-ion nuclear optical clock.
\end{abstract}

\section{Introduction}
\label{intro}
Optical atomic clocks have undergone fascinating development within the past two decades and are today approaching a time accuracy of $10^{-18}$ \cite{Ludlow}. They are based on a variety of different electronic transitions in the atomic shell and can be distinguished into two groups: single or multiple-ion atomic clocks and optical lattice clocks. In both clock concepts, a narrow-band laser is stabilized to an atomic transition and the number of laser oscillation periods is counted. After a certain number of oscillation periods, one second has elapsed. In a single-ion atomic clock the laser is stabilized to an atomic transition of a single ion stored and laser cooled in a Paul trap - examples of highest clock performance include $^{171}$Yb$^{+}$ \cite{Huntemann} and $^{27}$Al$^{+}$ \cite{Chen}. In an optical lattice clock, typically $10^4$ atoms are stored and irradiated in parallel to serve as a reference for laser stabilization - prominent examples include $^{87}$Sr \cite{Nicholson} and $^{171}$Yb \cite{Mcgrew}.\\[0.2cm]
In 2003, a new type of clock was proposed that uses a nuclear transition instead of an atomic shell transition for laser stabilization, which represents the main important difference to existing optical atomic clocks \cite{Peik}. This clock is typically referred to as the ``nuclear optical clock" (NOC) or simply ``nuclear clock". It is expected that the NOC will have some advantages compared to atomic shell-based clocks. The most intriguing one is that the nucleus is orders of magnitude smaller than the atom and, therefore, the transition frequency is significantly less affected by external pertubations. An achievable accuracy of a single-ion NOC of $1.5\cdot10^{-19}$ was estimated in 2012 \cite{Campbell}. Currently, two different approaches for a NOC are experimentally investigated, one based on individual ions stored in a Paul trap \cite{Peik,Campbell,Borisyuk1,Thielking} and the other one based on multiple ($\sim10^{15}$) ions in a solid-state crystal lattice environment \cite{Peik,Rellergert,Kazakov1,Jeet,Borisyuk,Stellmer}. The latter approach is comparable to an optical lattice clock, however, with the advantage of a drastically increased density of atoms.\\[0.2cm]
Direct nuclear laser excitation poses a central requirement for the development of a nuclear clock, however, this has not been experimentally achieved. In fact, only one nuclear transition is known to be accessible to modern laser technology, due to its extraordinary low excitation energy of about $7.8$~eV, corresponding to $\sim160$~nm wavelength \cite{Beck1,Beck2}. This is the first nuclear excited state in the $^{229}$Th isotope, which is metastable with a calculated radiative lifetime in the range of $10^4$~s \cite{Minkov,Tkalya} and therefore usually denoted as $^{229\mathrm{m}}$Th. Already in 1996, the $^{229}$Th nuclear isomeric transition has been identified to have properties useful for metrology \cite{Tkalya1}.\\[0.2cm]
An insufficient knowledge of the $^{229\mathrm{m}}$Th isomeric energy has so far hindered nuclear laser excitation of thorium ions and therefore the development of a NOC. This has led to a multitude of efforts to precisely determine the isomer's transition energy. A recent review is provided in Ref.~\cite{Wense1}. A particularly promising method is to use a high-resolution microcalorimetric $\gamma$-ray spectrometer similar to the measurement that has led to the currently best energy constraint \cite{Beck1,Beck2} in order to resolve a 29 keV $\gamma$-ray doublet, which is partly decaying to the ground and partly to the first excited state of $^{229}$Th \cite{Kazakov}. A corresponding experiment is currently conducted at the Kirchhoff-Institute for Physics in Heidelberg. A first time direct detection of the $^{229\mathrm{m}}$Th ground-state decay in 2016 \cite{Wense2} and a subsequent lifetime determination of the internal conversion process on a nickel alloy surface \cite{Seiferle1} have laid the foundation for three new ways of investigating the $^{229\mathrm{m}}$Th excitation energy. The first and most direct approach is to perform internal-conversion (IC) electron spectroscopy of the IC electrons emitted in the isomer's ground-state decay \cite{Seiferle2,Seiferle3}. A second approach is to use a superconducting nanowire single-photon detector (SNSPD) \cite{Natarajan} to investigate the isomeric energy based on a transition-edge detection technique and the third approach is to perform laser-based conversion-electron M\"ossbauer spectroscopy (CEMS) in a thin layer of neutral $^{229}$Th atoms \cite{Wense3}.\\[0.2cm]
The $^{229\mathrm{m}}$Th energy determination will most likely proceed in two steps: In a first step the energy will be determined by direct IC electron spectroscopy to about 100~meV precision \cite{Seiferle1}. Such measurement could be further confirmed by complementary techniques with a microcalorimetric $\gamma$-ray spectrometer or an SNSPD. In a second step direct laser spectroscopy on $^{229}$Th atoms can be performed, offering the potential to pin down the isomeric energy value by additional three orders of magnitude to 40~$\mu$eV, limited by the bandwidth of the laser system used for nuclear excitation, for which $\sim$10~GHz is assumed \cite{Wense3}. The energy would then be sufficiently constrained to allow for nuclear laser excitation of $^{229}$Th ions in a Paul trap and thus for the development of a single-ion nuclear clock.\\[0.2cm]
In the following we will focus only on the laser-based CEMS experiment, which is currently prepared in a collaboration of four groups located at the University of California, Los Angeles UCLA (laser development and scanning), the Leibniz University Hannover (laser development), the Johannes Gutenberg University Mainz ($^{229}$Th target fabrication) and Ludwig-Maximilians-University Munich (IC electron detection).
 
\section{The concept of laser-based CEMS}
\label{sec:1}
The concept of laser-based conversion-electron M\"ossbauer spectroscopy (CEMS) was first proposed in 2017 for direct nuclear laser excitation of $^{229\mathrm{m}}$Th \cite{Wense3}. The central idea of this concept is to perform nuclear laser spectroscopy of neutral $^{229}$Th atoms instead of ions, as usually considered. This has the advantage that the internal conversion (IC) decay channel is not suppressed, thereby reducing the isomeric lifetime to about 10~$\mu$s \cite{Tkalya,Seiferle1,Karpeshin}. This allows probing for successful $^{229\mathrm{m}}$Th laser excitation after each individual laser pulse, when a pulsed laser system is used for scanning. Whether the isomer was successfully excited is determined by detecting the IC electrons that will subsequently be emitted in the isomeric decay following each individual laser pulse. The IC electron detection can then be correlated with the laser pulses, which drastically improves the achievable signal-to-background ratio. A sketch of the experimental concept is shown in Fig.~1.
\end{multicols}
\begin{figure}[h]
\includegraphics[width=0.9\textwidth]{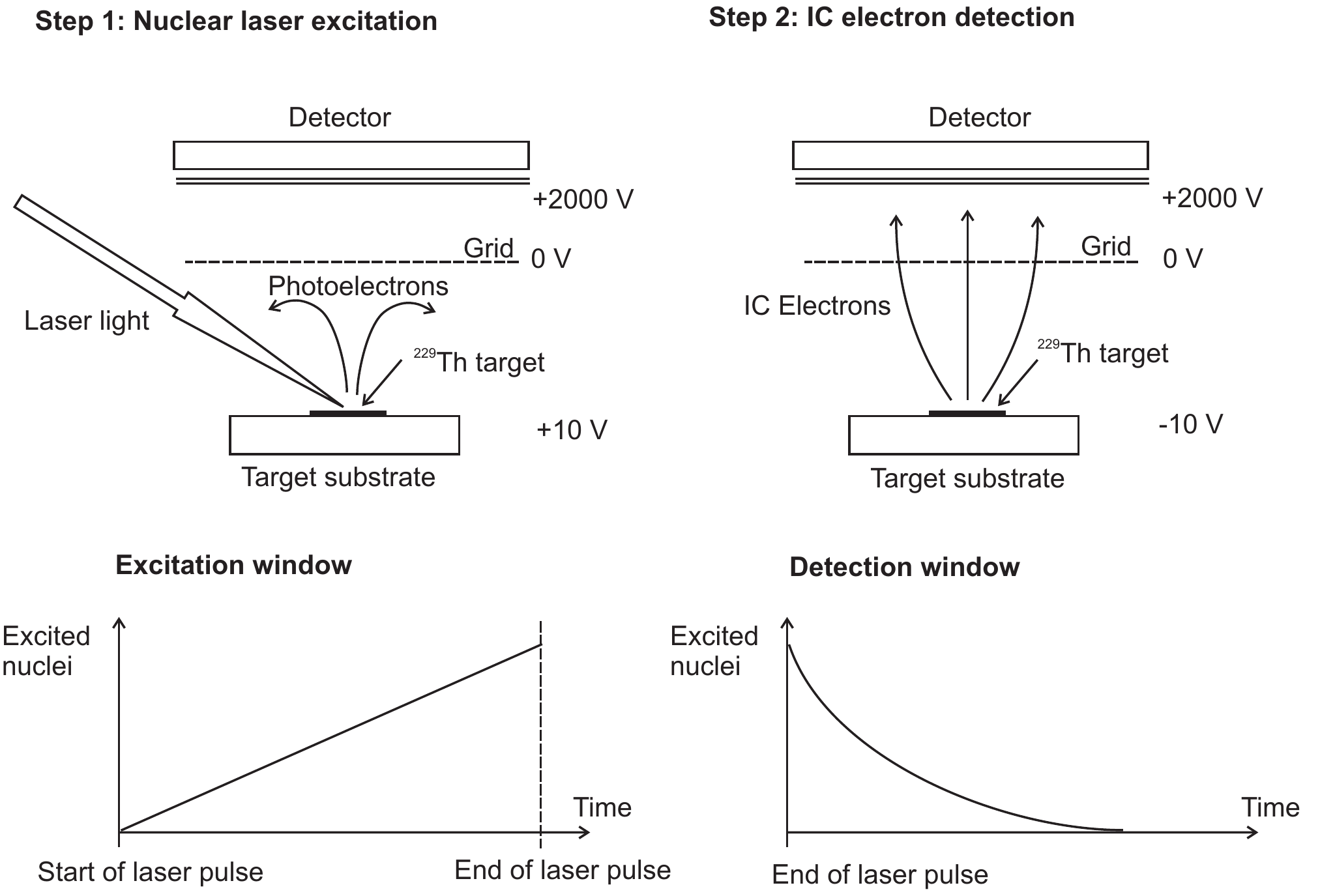}
\caption{Conceptual sketch of the nuclear laser excitation scheme. The concept is divided into two steps: In the first step the $^{229}$Th nuclei deposited on a surface are laser excited. No detection of isomeric decays is possible during this time due to the strong background from photoelectrons. In the second step the laser excitation is terminated and electrons emitted in the isomer's internal conversion (IC) decay are observed. Between the two steps, the target voltage is changed: During step 1, photoelectrons emitted from the target are hindered from reaching the MCP detector in order to prevent detector saturation during laser irradiation. During step 2 the target electrons can reach the MCP detector to allow for IC electron detection.}
\label{fig:1}   
\end{figure}
\begin{multicols}{2}

\subsection{The $^{229}$Th target}
\label{sec:2}
Using neutral $^{229}$Th atoms as a target allows for high atomic densities and thus the ability to irradiate a large number of nuclei in parallel. In an ideal scenario, a thin, metallic $^{229}$Th layer would be used as a target in order to achieve the highest density, although, for practical reasons, some chemical compound, like ThO$_2$, might be more realistic. 
For the following estimates we assume that a $^{229}$ThO$_2$ layer with a density of 9.86 gcm$^{-3}$ is laser irradiated.\\[0.2cm]
The amount of material should be kept at the minimum required for the experiment, as $^{229}$Th is an $\alpha$ emitter with about 7920 years half-life \cite{Varga} and the radioactive decays of $^{229}$Th and its daughters lead to an unavoidable background. The layer thickness should not exceed the optical penetration depth of the VUV laser light in the material (estimated to 11.6~nm \cite{Wense3}) and has to be sufficiently thin in order to also allow for the conversion electrons to efficiently leave the surface. The mean free path length of electrons 7.8~eV above the Fermi level in solids can be estimated to be $\sim3$~nm \cite{Seah}. However, as most of the electrons in the IC decay will be emitted from electronic levels below the Fermi level, the kinetic energy of the majority of electrons in the material will be below 7.8~eV. As the mean free path increases to 10~nm for electrons at 3.4~eV (corresponding to the thorium work function \cite{Reviere}) above the Fermi level, a target thickness of about 10~nm appears to be a reasonable choice. As thorium atoms in ThO$_2$ occupy sites on an fcc lattice with a lattice constant of about 560 pm, this corresponds to a few tens of thorium atomic layers.\\[0.2cm]
The target area should not significantly exceed the area of the laser spot, for which a diameter of 1~mm is assumed in the following. A straight forward calculation reveals, that a 10~nm thick layer of $^{229}$ThO$_2$ with 1~mm diameter contains about $1.8\cdot10^{14}$ atoms. This corresponds to an intrinsic $^{229}$Th activity of 500~Bq, which can easily be handled. Customized $^{229}$Th targets designed for this experiment are currently prepared at the Institute of Nuclear Chemistry of the University of Mainz via a newly developed drop-on-demand technique \cite{Haas}. The envisaged target parameters are listed in Tab.~\ref{tab:1}.
\begin{table}[H]
\caption{Parameters of the $^{229}$Th target}
\label{tab:1}
\begin{tabular}{ll}
\hline\noalign{\smallskip}
Parameter  & Value  \\
\noalign{\smallskip}\hline\noalign{\smallskip}
Diameter  & 1 mm \\
Thickness  & 10 nm\\
Density  & 9.86 gcm$^{-3}$\\
Number of atoms & $1.8\cdot10^{14}$\\
Activity &  500~Bq\\
\noalign{\smallskip}\hline
\end{tabular}
\end{table}

\subsection{The laser system}
\label{sec:3}
\begin{figure}[H]
\includegraphics[width=0.5\textwidth]{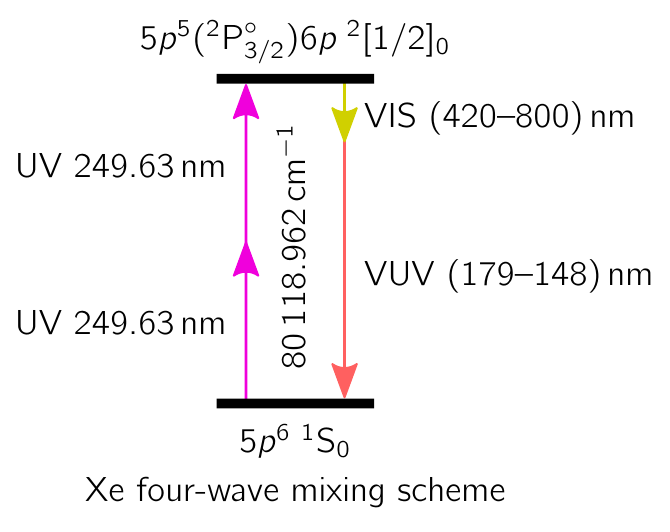}
\caption{Resonance-enhanced four-wave mixing scheme in Xe gas. The $5p^5(^2\mathrm{P}^\circ_{3/2})6p\;{}^{2}[1/2]_0$ level in neutral xenon is excited from the ground state via a two-photon excitation induced by PDL~1. The deexcitation can occur also via two photons, where the frequency of the first photon is given by tunable PDL~2.}
\label{fourwavescheme}   
\end{figure}

\begin{figure}[H]
\includegraphics[width=1.0\textwidth]{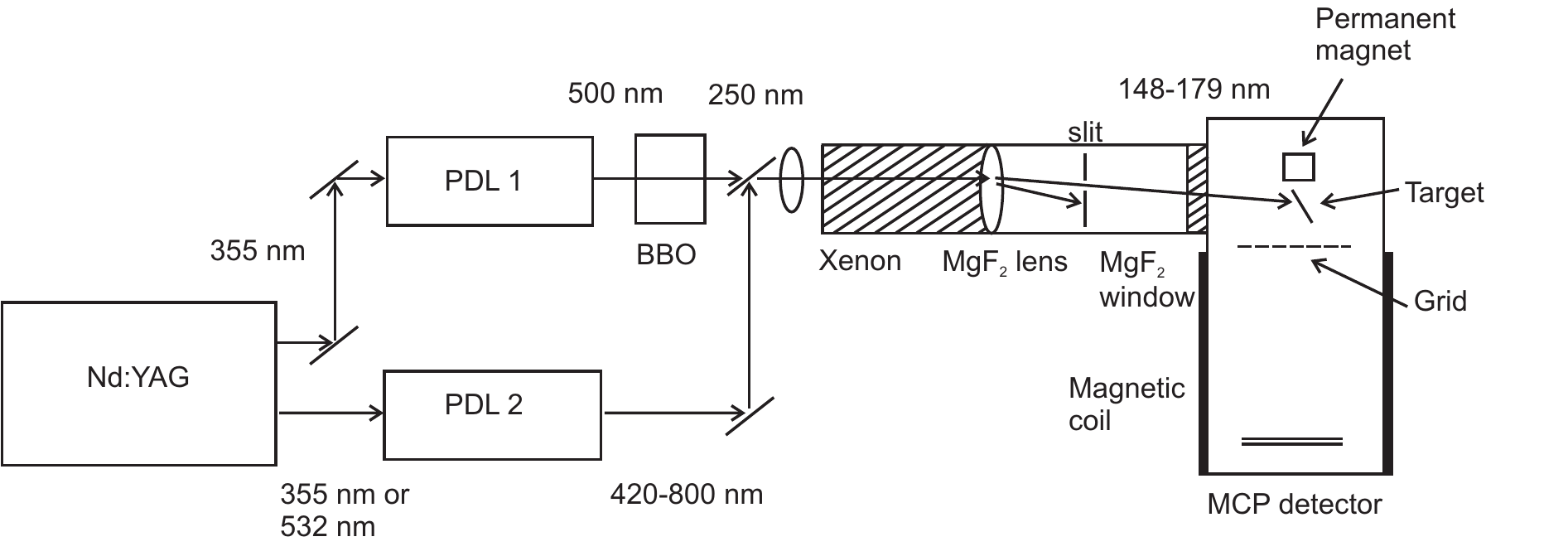}
\caption{Schematic drawing of the experimental setup. A tunable and pulsed VUV laser beam is generated via resonance enhanced four-wave mixing in a xenon gas. Two pulsed dye lasers (PDLs) are used for this purpose. The laser light is used to irradiate a thin (about 10~nm thick) $^{229}$Th layer, which is placed inside of a magnetic bottle generated by a strong (1 Tesla) permanent magnet and a magnetic coil. In case of successful nuclear laser excitation, low energy electrons, that are emitted in the internal conversion decay of $^{229\mathrm{m}}$Th, will be able to leave the $^{229}$Th target and will be guided by the magnetic fields towards a micro-channel-plate (MCP) detector.}
\label{fig:2}   
\end{figure}

From the described concept, it is obvious that a pulsed laser system is required. The pulse duration should not be significantly longer than the IC-shortened isomeric lifetime, as otherwise the excited state population would already be in equilibrium and any further irradiation would not increase the signal. Further, the pulse duration should not be too short, as in this case the increased pulse power could lead to ablation or even evaporation of the target, which is unwanted. The typical metal ablation threshold is about $10^8$~W/cm$^{2}$ \cite{Zimmermann}. Thus the useful laser energy per pulse is limited and the maximum allowed value depends on the pulse duration. On the other hand, the pulse energy has to be sufficiently large to lead to a detectable amount of nuclear excitations per pulse. An important further requirement is a broad tunability in the vacuum ultra-violet (VUV) spectral region, as the energy of the isomeric state has so far been only roughly constrained. The laser bandwidth should not be too narrow, in order to limit the number of required steps for scanning.\\[0.2cm]
A laser system that fulfills all the above requirements and can therefore be used for the experiment was reported, e.g., in Ref.~\cite{Hanna}. A similar system is now operational at UCLA and has already been used for laser irradiation of $^{229}$Th-doped crystals. The system is based on two pulsed dye lasers (PDLs), both pumped by an injection-seeded Nd:YAG laser with up to 1.3~J/pulse at 1064~nm and a repetition rate of 30 Hz. The first PDL operates at a fixed wavelength of $\sim500$~nm, which is then frequency doubled in a BBO (Beta Barium Borate) crystal to $\sim250$ nm ($\omega_1\approx7.5$~PHz), resonant to the two-photon transition between ground state and $5p^5(^2\mathrm{P}^\circ_{3/2})6p\;{}^{2}[1/2]_0$ state in neutral xenon (at $\sim 125$~nm). The second PDL delivers tunable light between 420 and 800~nm ($\omega_2\approx 2.4$~PHz to $4.5$~PHz). Both laser beams are then mixed via resonance-enhanced four-wave mixing in a cell containing xenon gas (see Fig. \ref{fourwavescheme} for the four-wave mixing scheme). Using the difference component $\omega_3=2\omega_1-\omega_2$, we obtain a source of pulsed laser light, broadly tunable between $\omega_3=10.5$ and $12.6$~PHz, corresponding to a covered energy range between $6.9$ and $8.3$~eV. Extending the VUV tuning range to at least 15.9 eV is straightforward by selecting the sum frequency or using different resonances in xenon or krypton in combination with a free gas jet instead of a gas cell \cite{Ng}. A conceptual sketch of the experimental setup is shown in Fig.~\ref{fig:2}.
\end{multicols}
\begin{multicols}{2}
\noindent The laser system can produce up to 40~$\mu$J/pulse in the VUV and about 10~$\mu$J/pulse can be maintained over most of the tuning range. 
The bandwidth ($\Gamma_L$) of the laser light is calculated from the PDL's envelope linewidths ($\Gamma_\mathrm{PDL1}\approx3$~GHz for the UV PDL and $\Gamma_\mathrm{PDL2}\approx3$ to $6$~GHz for the visible PDL) via $\Gamma_L=\sqrt{4\Gamma_\mathrm{PDL1}^2+\Gamma_\mathrm{PDL2}^2}$ to be below $6.7$~GHz. A value of 10~GHz at a pulse duration of 10~ns is assumed for all following calculations.\\[0.2cm]
The PDLs' bandwidth is the envelope of many (about 20) PDL cavity modes with a minimum of 50 MHz bandwidth, which are spaced by a frequency of 450~MHz \cite{Kajava}. The cavity end mirror of PDL~2 is dithered by a bit more than the free spectral range within 1 second to smear out the cavity spectrum over the course of the illumination. (For this reason, in case of resonance, an isomeric decay signal would occur in about three out of the 30 pulses per second with a 1 second period and a signal rate which is by a factor of about 10 enhanced compared to the predicted pulse-averaged signal rate given in section~\ref{sec:5}).\\[0.2cm]
These light pulses are then used to irradiate the $^{229}$Th target surface of 1~mm diameter. A straight forward calculation leads to a pulse intensity of $1.3\cdot10^5$ W/cm$^2$, which is three orders of magnitude below the laser ablation threshold. The most important laser parameters are listed in Tab.~\ref{tab:2}. An alternative setup, based on a broadly tunable Ti:Sapphire laser system is currently prepared at the Leibniz University in Hannover.
\begin{table}[H]
\caption{Parameters of the UCLA tunable VUV laser system}
\label{tab:2}
\begin{tabular}{ll}
\hline\noalign{\smallskip}
Parameter & Value  \\
\noalign{\smallskip}\hline\noalign{\smallskip}
Pulse energy  & 10 $\mu$J \\
Pulse duration  & 10 ns\\
Pulse power  & $10^3$ W\\
Spot diameter  & 1 mm  \\
Pulse intensity  & $1.3\cdot10^5$ W/cm$^2$\\
Bandwidth  & $2\pi\cdot10$ GHz\\
Rep. rate  & 30 Hz \\
\noalign{\smallskip}\hline
\end{tabular}
\end{table}

\subsection{The detection system}
\label{sec:4}
\begin{center}
\begin{figure}[H]
\includegraphics[width=1.0\textwidth]{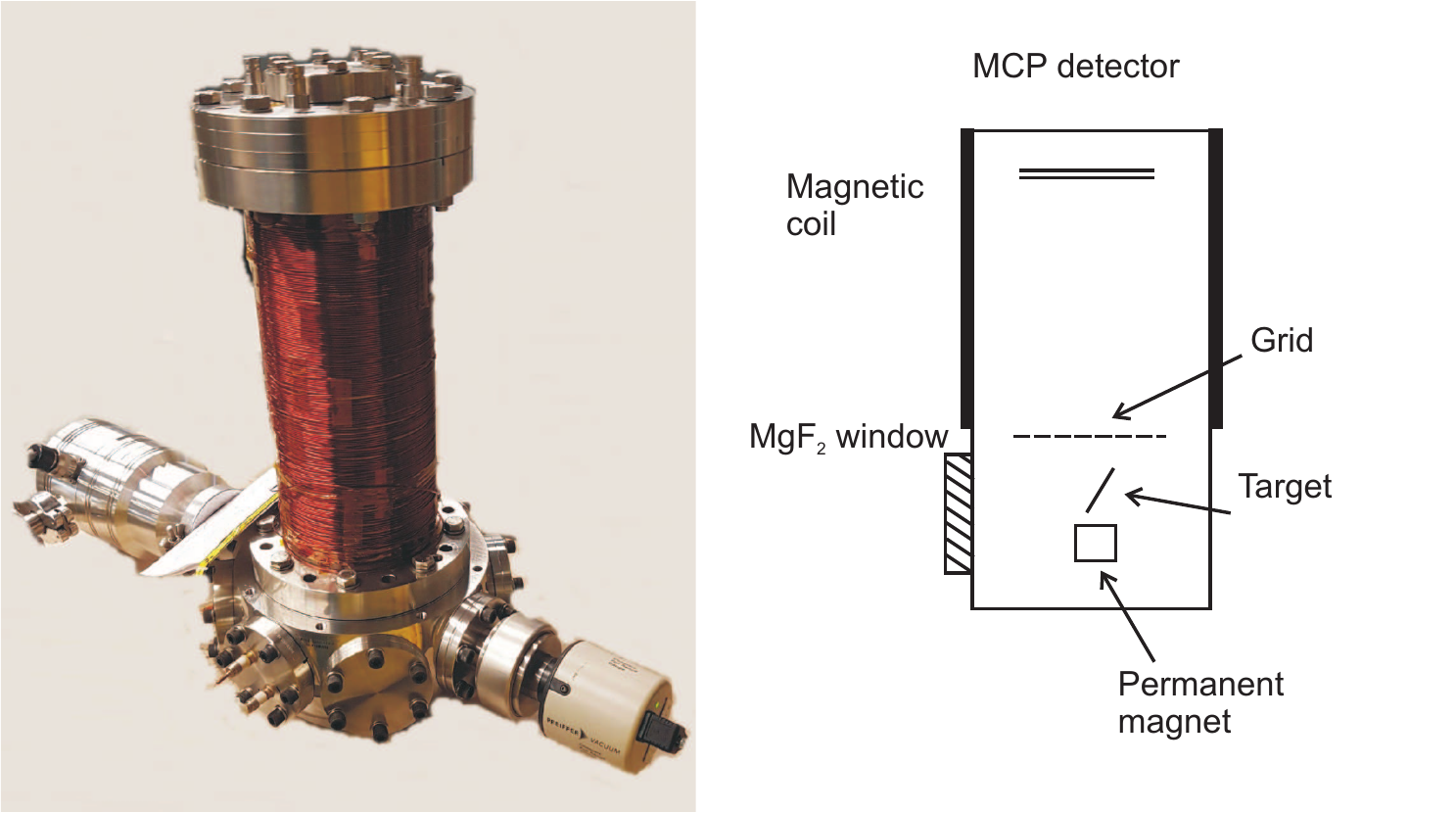}
\caption{Photograph of the vacuum chamber used for IC-decay detection. The magnetic coil used for electron guiding is placed outside of the vacuum chamber to allow for UHV conditions. The distance between the $^{229}$Th target and the MCP detector is about 30~cm. This will reduce the background originating from high energy particles as generated in the radioactive decay processes within the target.}
\label{fig:3}   
\end{figure}
\end{center}

During an individual laser pulse, a certain fraction of nuclei will be excited into the isomeric state if the laser is tuned to the resonance frequency of the nucleus. The excited nuclei will decay into their ground states via internal conversion through the emission of an electron. As the target layer is thin, a significant fraction of these electrons will be able to leave the sample surface. As mentioned before, typical mean free path lengths of the low energy electrons range between 2.5 and 10~nm \cite{Seah} and the design value of the target layer thickness is 10~nm. Ultra-high-vacuum (UHV) conditions are foreseen in order to reduce a potential carbon layer ongrowth induced by the VUV irradiation. The electrons will be guided by magnetic fields that are generated by a strong ($\sim$1~T) permanent magnet in combination with a magnetic coil, before being detected by a micro-channel plate (MCP) detector, providing a detection efficiency of up to 50\% if the electrons are post-accelerated to above 300~eV kinetic energy \cite{Goruganthu}. Due to the magnetic guiding fields, mostly low energy electrons are detected and background in the form of high energy particles emitted during the radioactive decays in the target will be suppressed. Nevertheless, we expect the background to the isomeric decay signal to be dominated by low-energy electrons that are generated in the radioactive decay processes. Background signals generated due to delayed photoelectrons triggered by the laser irradiation will fade away within a few microseconds after the end of the laser pulse (see sect.~\ref{sec:6}). A retarding grid is placed between the target and the detector and the target voltage can be set to a positive potential. In this way photoelectrons are hindered from reaching the MCP detector and saturation of the detector due to excessive photoelectron bombardment during laser irradiation is prevented \cite{Arlt}. A certain time after the end of the laser pulse the target potential is switched to a negative value, thereby allowing electrons emitted from the sample to reach the detector.\\[0.2cm] 
Preparatory measurements were performed in order to estimate the strength of background originating from delayed photoelectrons. These measurements will be described in section~\ref{sec:6}. The result is that photoelectron background can be expected to only play a minor role and will not prevent the experimental concept. A photograph of the vacuum chamber designed for the IC electron detection is shown in Fig.~\ref{fig:3}.
\end{multicols}
\begin{multicols}{2}
\section{Quantitative analysis of the concept}
\label{sec:5}
There are three central questions to be answered in order to judge the validity of the proposed laser excitation scheme: (1) What is the expected absolute number of nuclear excitations per time? (2) What is the expected signal-to-background ratio? (3) How long will it take to scan the large energy range of 1~eV or more, corresponding to the uncertainty of the isomeric energy value? These questions will be addressed in this section.\\[0.2cm]

\subsection{What is the expected absolute number of nuclear excitations?}
In Refs.~\cite{Wense3,Wense4} it was shown that the number of excited nuclei $N_\mathrm{exc}$ of a nuclear two-level system, when resonantly irradiated with a broad-band and low-intensity source of light, can be calculated as a function of excitation time $t_\mathrm{exc}$ via
$$
N_\mathrm{exc}(t_\mathrm{exc})=\frac{g_\mathrm{exc}}{g_\mathrm{gnd}}\frac{\rho^\omega\pi^2c^3\Gamma_\gamma N_0}{\Gamma_\mathrm{tot}\hbar\omega^3}\left(1-\mathrm{e}^{-\Gamma_\mathrm{tot}t_\mathrm{exc}}\right).
$$
Here $\rho^\omega$ denotes the spectral energy density, which is best approximated as $\rho^\omega=2I/(\pi c\Gamma_L)$, with $I$ the intensity and $\Gamma_L$ the bandwidth of the laser light used for irradiation. $N_0$ is the number of irradiated nuclei, $\Gamma_\mathrm{tot}=(1+\alpha_\mathrm{ic})\Gamma_\gamma$ the total decay rate of the excited nuclear state including both, $\gamma$ decay and internal conversion (IC), with $\alpha_\mathrm{ic}$ as the IC coefficient and $\Gamma_\gamma$ the radiative decay rate. $\omega$ denotes the angular frequency corresponding to the nuclear excitation. $g_\mathrm{gnd}=2j_\mathrm{gnd}+1$ and $g_\mathrm{exc}=2j_\mathrm{exc}+1$ are the degeneracies of the ground and excited state, respectively, where $j_\mathrm{gnd}=5/2$ and $j_\mathrm{exc}=3/2$ are the angular momentum quantum numbers in case of $^{229}$Th. For the parameters listed in Tab.~\ref{tab:3} a calculation reveals that the number of excited nuclei per laser pulse amounts to $8.0\cdot10^3$.
\begin{table}[H]
\caption{Variables and values used for calculating the number of nuclear excitations per laser pulse.}
\label{tab:3}
\begin{tabular}{lll}
\hline\noalign{\smallskip}
Parameter & Variable & Value  \\
\noalign{\smallskip}\hline\noalign{\smallskip}
Pulse intensity & $I$ & $1.3\cdot10^5$ W/cm$^2$ \\
Laser bandwidth & $\Gamma_L$ & $2\pi\cdot10$ GHz\\
Spectr. energy dens. & $\rho^\omega$ & $4.4\cdot10^{-11}$ J/(m$^{3}$Hz)\\
Rad. decay rate & $\Gamma_\gamma$ & $10^{-4}$ Hz\\
Number of atoms & $N_0$ & $1.8\cdot10^{14}$\\
IC coefficient & $\alpha$ & $10^9$\\
Total nucl. dec. rate & $\Gamma_\mathrm{tot}$ & $10^5$ Hz\\
Ang. trans. frequ. & $\omega$ & $2\pi\cdot1.9$ PHz\\
Pulse time & $t_\mathrm{exc}$ & $10$ ns\\
Gnd. state degen. & $g_\mathrm{gnd}$ & $6$ \\ 
Exc. state degen. & $g_\mathrm{exc}$ & $4$ \\ 
Number of exc. nucl. & $N_\mathrm{exc}$ & $8.0\cdot10^3$ \\ 
\noalign{\smallskip}\hline
\end{tabular}
\end{table}
\noindent These nuclei will decay immediately after the laser pulse with an expected lifetime of $\tau\approx10$~$\mu$s, following the exponential decay curve given by $N_\mathrm{dec}(t_\mathrm{dec})=N_\mathrm{exc}\mathrm{e}^{-t_\mathrm{dec}/\tau}$. In order to not contaminate the signal with delayed photo-electrons, the electron detection will start shortly (about 5~$\mu$s) after the laser pulse. The number of occurring IC electrons $N_\mathrm{ic}$ equals the number of isomeric decays in the time interval used for detection. Therefore we have
$$
\begin{aligned}
N_\mathrm{ic}&=-\int_{t_\mathrm{start}}^{t_\mathrm{end}}\frac{dN_\mathrm{dec}}{dt_\mathrm{dec}}\ dt_\mathrm{dec}=\left[-N_\mathrm{exc}\mathrm{e}^{-\frac{t_\mathrm{dec}}{\tau}}\right]_{t_\mathrm{start}}^{t_\mathrm{end}}\\
&=N_\mathrm{exc}\left[\mathrm{e}^{-\frac{t_\mathrm{start}}{\tau}}-\mathrm{e}^{-\frac{t_\mathrm{end}}{\tau}}\right].\\
\end{aligned}
$$
Assuming values of $t_\mathrm{start}=5$~$\mu$s and $t_\mathrm{end}=10$~$\mu$s, one obtains $N_\mathrm{ic}=N_\mathrm{exc}\left[\mathrm{e}^{-1/2}-\mathrm{e}^{-1}\right]=0.24\cdot N_\mathrm{exc}$. Thus the number of emitted IC electrons in the 5~$\mu$s time window can be estimated to $1800$ per laser pulse. With a pulse repetition rate of 30~Hz, this leads to a total rate of $54,000$ electrons per second. Further losses may arise due to a limited optical penetration depth of the light in the material, electron losses in the material and the guiding and detection efficiency. These effects may easily lead to a factor of 10 decrease in absolute count rate, still resulting in an expected number of detected IC electrons of $180$ per laser pulse in a 5~$\mu$s time window.

\subsection{What is the expected signal-to-background ratio?}
Any estimate of the expected signal-to-background ratio requires assumptions on the expected background. We anticipate that two effects will dominate the background, which are (1) low-energy electrons generated by radioactive decays in the $^{229}$Th target and (2) delayed photoelectrons triggered by the laser irradiation. The first type of background, although occurring with equal time distribution and therefore to a large extent suppressed due to the time gating, is expected to dominate the background. The number of radioactive $^{229}$Th decays in the target was estimated to 500 per second. This number, however, will grow soon after chemical purification by a factor of about 10, due to the ingrowth of short-lived daughter nuclei. As each radioactive decay will on average lead to the emission of two low-energy electrons \cite{Wandkowski}, we estimate the total emission rate of low energy electrons of the target to $1\cdot10^4$ per second. As these electrons will be equally distributed in time, the number of detected electrons in a 5~$\mu$s time window amounts to $5\cdot10^{-2}$. This compares to about $180$ IC electrons in the same time window, resulting in a signal-to-background ratio of $3600:1$.\\[0.2cm]
The delayed photoelectrons, although occurring time correlated with the laser pulses and therefore having the potential to pose a serious problem for the proposed experiment, could experimentally be shown to be of minor concern. In order prove this, we have performed preparatory experiments described in section~\ref{sec:6}. As a result, an upper limit for the number of delayed photo-electrons occurring in the time window between 5 and 10~$\mu$s after the end of each laser pulse can be given as $2.5\cdot10^{-2}$ per pulse. Adding this to the background generated by radioactive decays lead to $7.5\cdot10^{-2}$ electrons per pulse, which compares to 180 signal electrons in the same time window, leading to a total expected signal-to-background ratio of $2400:1$.

\subsection{How long will it take to scan the large energy range of 1~eV?}
The time required to scan the large energy range of 1~eV is an important parameter, as this is currently the uncertainty range of the isomeric energy, which has been constrained to $(7.8\pm0.5)$~eV \cite{Beck1,Beck2}. The 10~GHz bandwidth of the laser used for scanning corresponds to $40$~$\mu$eV. Therefore the number of required scan steps amounts to $25,000$. Assuming that 300 laser pulses (corresponding to 10~s irradiation time) are used for each individual scan step, the total time required to scan 1~eV in energy amounts to $2.5\cdot10^5$~s or 2.9 days. This does not include laser maintenance, dye changes, etc. In case of resonance, the number of detected IC electrons would amount to $54,000$ in the 300 time windows of 5~$\mu$s duration.

\section{Preparatory experiments}
\label{sec:6}
\begin{center}
\begin{figure}[H]
\includegraphics[width=0.7\textwidth]{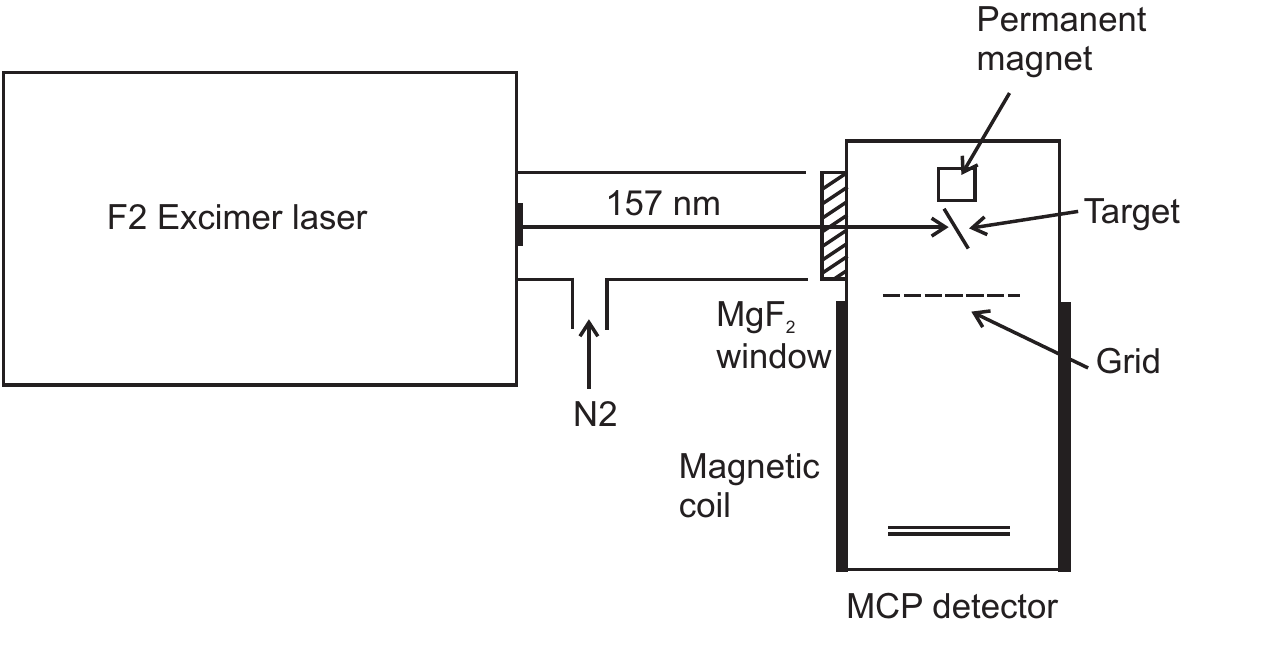}
\caption{Schematic drawing of the experimental setup used to investigate the laser-induced electronic background. The tunable VUV laser source of Fig.~\ref{fig:2} is exchanged by an F$_2$ excimer laser providing a fixed wavelength of 157~nm. The laser is coupled to the vacuum detection chamber with a constant nitrogen flow to prevent intensity reduction by absorption due to the oxygen in air.}
\label{fig:4}   
\end{figure}
\end{center}

\begin{center}
\begin{figure}[H]
\includegraphics[width=0.5\textwidth]{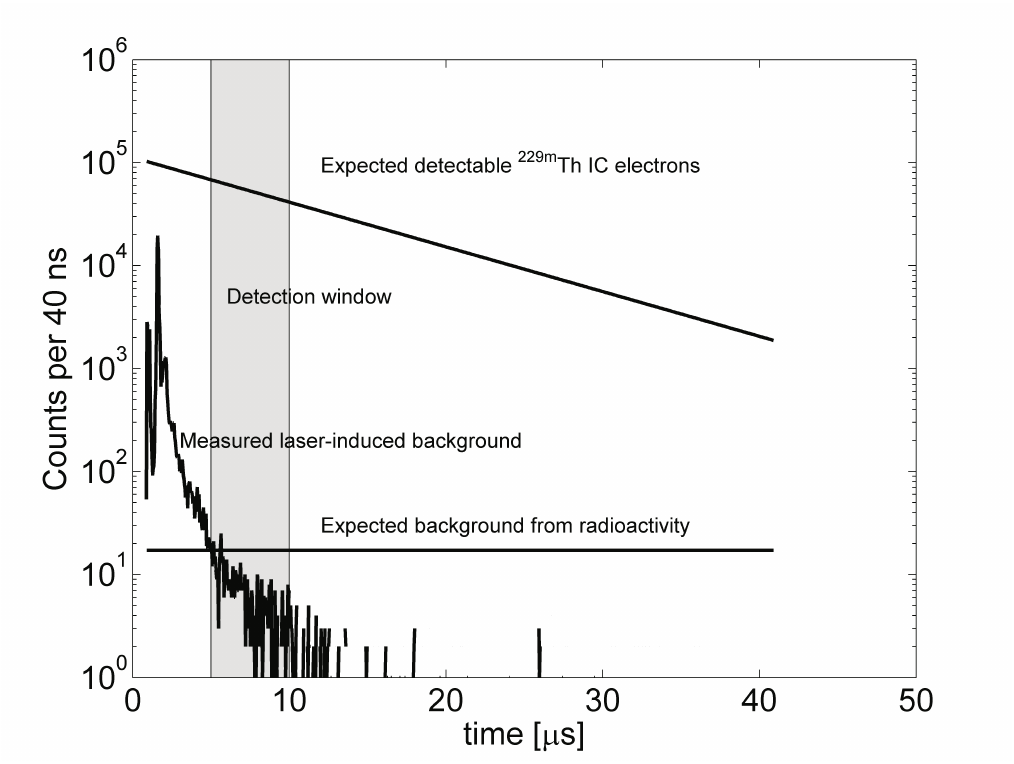}
\caption{Experimental results of the preparatory measurement. The time resolved number of delayed photoelectrons is shown for a total of $3.3\cdot10^4$ laser pulses. The expected number of radioactivity-induced background electrons as well as internal conversion electrons in case of resonance under identical measurement conditions are also shown. In the time window from 5 to 10~$\mu$s the total number of detected photoelectrons amounts to $831$. The expected number of detected IC electrons in the same time interval is $5.9\cdot10^6$ ($180\cdot3.3\cdot10^4$), leading to an expected signal-to-background ratio of $7000:1$. This estimate assumes a factor of 10 loss of IC electrons due to limited guiding and detection efficiencies. The figure was adapted from Ref.~\cite{Arlt}.}
\label{fig:5}   
\end{figure}
\end{center}

Laser-induced delayed electrons could introduce a background signal sufficiently large to prevent the IC electrons from being detectable. In order to investigate this question, we used our IC spectroscopy setup and irradiated a metallic (aluminum) target with a pulsed VUV laser source. As tunability of the laser was not a requirement for this test measurement, an F$_2$ excimer laser (GAM Laser EX5F) operating at a fixed wavelength of 157~nm was used. The laser provides pulses of 8~ns pulse duration with an energy of 2 mJ per pulse and a beam profile of $3\times 6$~mm$^2$. A repetition rate of 20~Hz was used in the experiment. The pulse intensity is inferred to be $1.4\cdot10^6$ W/cm$^2$, which is an order of magnitude larger than the pulse intensity of the laser that will be used for $^{229}$Th laser excitation. For this reason we consider the results obtained in this preparatory experiment to represent an upper limit on the expected laser-induced electron background. A sketch of the experimental setup is shown in Fig.~\ref{fig:4}.
\end{multicols}
\begin{multicols}{2}
\noindent The excimer laser is coupled to the target chamber via a MgF$_2$ window and a tube providing a nitrogen flow. The nitrogen flow is required to sufficiently reduce the amount of oxygen, which would otherwise absorb the VUV light.\\[0.2cm]
The result of the measurement is shown in Fig.~\ref{fig:5} \cite{Arlt} together with the expected number of detectable IC electrons ($10\%$ of the total number of isomeric decays) and the expected radioactivity-induced background for the same conditions. A total number of $\sim3.3\cdot10^4$ laser pulses were used for this measurement. In the time window between 5 and 10~$\mu$s the number of detected delayed background electrons amounts to $831$, which corresponds to $0.025$ electrons per laser pulse. It can be inferred from the measurement that, from the point of laser-induced signal-to-background ratio, it might even be advantageous to choose a later time window for the detection. However, already at this point it can be expected that the background will be dominated by the target radioactivity instead of the photoelectrons.

\section{Summary and perspectives}
The described method of laser-based conversion-electron M\"ossbauer spectroscopy (CEMS) offers potential for a first nuclear laser spectroscopy. The method is several orders of magnitude superior compared to other potential direct nuclear laser excitation schemes in terms of signal-to-background ratio and measurement time. In case of success, the method could be used to constrain the isomeric energy to 40~$\mu$eV, corresponding to the 10~GHz bandwidth of the laser used for excitation. This is an improvement by a factor of $2.5\cdot10^4$ compared to the currently best value of $(7.8\pm0.5)$~eV. The precision could be even further narrowed down if a laser providing a smaller bandwidth would be used in a second step. The achievable precision is sufficient to allow for direct nuclear laser excitation of trapped thorium ions and thus for a doorway into the development of a nuclear clock of ultra-high accuracy. In Ref.~\cite{Wense1} it was emphasized that laser-based CEMS could also be used in a clock concept. In this case the linewidth of the nuclear transition would be IC-broadened to 15.9~kHz, which is comparable to the expected broadening of the nuclear transition in the crystal-clock approach of up to 10~kHz \cite{Rellergert}. A nuclear optical clock, when operational, is expected to impact on time metrology \cite{Peik,Campbell} as well as on relativistic geodesy \cite{Mehlstaeubler} and act as a probe for potential time variations of fundamental constants \cite{Peik,Rellergert,Flambaum}.

\begin{acknowledgements}
We would like to thank S. Stellmer and T. Schumm for discussions and lending of the VUV excimer laser. For discussions we are also grateful to G. Kazakov, A. P\'{a}lffy, J. Weitenberg and E. Peik. This work was supported by DFG (Th956/3-2) and by the European Union's Horizon 2020 research and innovation programme under grant agreement 6674732 "nuClock". The efforts at UCLA have been supported in part by DARPA (QuASAR program), ARO (W911NF-11-1-0369), NSF (PHY-1205311), NIST PMG (60NANB14D302), RCSA (20112810), and DOE Office of Nuclear Physics, Isotope Programme.
\end{acknowledgements}



\end{multicols}
\end{document}